# An Efficient Spectral Leakage Filtering for IEEE 802.11af in TV White Space

Phu Xuan Nguyen[1], Thinh Hung Pham[2], Trang Hoang[2], and Oh-Soon Shin[3*]

[1]School of Electronic Engineering, Soongsil University, Korea
[2]Department of Electrical – Electronic Engineering, Ho Chi Minh City University of Technology, Vietnam
[3]School of Electronic Engineering and Department of ICMC Convergence Technology, Soongsil University, Korea
* Corresponding author (E-mail: osshin@ssu.ac.kr)

*Abstract*—Orthogonal frequency division multiplexing (OFDM) has been widely adopted for modern wireless standards and become a key enabling technology for cognitive radios. However, one of its main drawbacks is significant spectral leakage due to the accumulation of multiple *sinc*-shaped subcarriers. In this paper, we present a novel pulse shaping scheme for efficient spectral leakage suppression in OFDM based physical layer of IEEE 802.11af standard. With conventional pulse shaping filters such as a raised-cosine filter, vestigial symmetry can be used to reduce spectral leakage very effectively. However, these pulse shaping filters require long guard interval, i.e., cyclic prefix in an OFDM system, to avoid inter-symbol interference (ISI), resulting in a loss of spectral efficiency. The proposed pulse shaping method based on asymmetric pulse shaping achieves better spectral leakage suppression and decreases ISI caused by filtering as compared to conventional pulse shaping filters.

## I. INTRODUCTION

Orthogonal frequency division multiplexing (OFDM) is being widely used for many wireless communication systems as well as Cognitive Radios (CR) due to many advantages [1]. OFDM divides a wideband channel into narrowband flat-fading subchannels. Inter-symbol interference (ISI) caused by multipath propagation can be eliminated by inserting a cyclic prefix (CP) into each OFDM symbol. Moreover, OFDM makes efficient use of spectrum by allowing overlapping of subcarriers. However, OFDM has several drawbacks, one of which is spectral leakage caused by the accumulation of multiple *sync*-shaped subcarriers [2]. This problem can be resolved by the use of a well-designed pulse shaping filter.

To reduce spectral leakage, it is necessary to increase the roll-off factor of the pulse shaping filter. Conventional pulse shaping filters restrict the roll-off region to be smaller than the length of the CP to prevent ISI. This implies that an increase in the roll-off factor necessitates an increase in the CP length, deteriorating the spectral efficiency [3], [4], [5]. Recently, development of 5G system requires an OFDM system with high spectral efficiency without increasing the CP overhead. Asymmetric pulse shaping filter was introduced to achieve such a goal. We propose the use of an asymmetric pulse shaping filter for IEEE 802.11af standard, which is based on OFDM and operates in TV white space (TVWS). Note that TVWS is a portion of the spectrum not used by TV broadcasting in VHF and UHF terrestrial TV bands [6]. IEEE 802.11af is a global standard for wireless local area network (WLAN) that uses CR technology to operate in the TVWS [7]. The PHY layer configuration of the IEEE 802.11af is defined based on the VHT (very high throughput) 40 MHz mode of the IEEE 802.11ac [6], [8]. One problem to be resolved is to obtain the 55dB attenuation requirement of its spectrum emission mask (SEM) [7].

In this paper, we present a novel pulse shaping scheme for efficient spectral leakage suppression in OFDM based physical layer of IEEE 802.11af. The proposed pulse shaping method is based on asymmetric pulse shaping and extension of guard band. To the best of our knowledge, there is no published filtering method that can suppress spectral leakage of 802.11af signal with a short guard interval. The main contributions of this paper are summarized as follows.

- The paper presents a novel filtering scheme for 802.11af systems that use a short guard interval, i.e., CP, not to degrade the spectral efficiency. Using the method, the spectrum leakage of the IEEE 802.11af can be low enough to meet the SEM.
- The use of an asymmetric pulse shaping can overcome the limited smoothing duration constrained by a short CP.
- Extending frequency guard interpolation by increasing the number of FFT is presented. This significantly limits the effect of image spectra, leading to reduced FIR filter length.

The rest of this paper is organized as follows. Section II describes an OFDM system model and introduces IEEE 802.11af standard. In Section III, we present the proposed pulse shaping scheme. Numerical results are presented and discussed in Section IV. Finally, conclusions are drawn in Section V.

This work was supported in part by Institute for Information & Communications Technology Promotion (IITP) grant funded by the Korean government (MSIT) (No. 2017-0-00724, Development of Beyond 5G Mobile Communication Technologies (Ultra-Reliable, Low-Latency, and Massive Connectivity) and Combined Access Technologies for Cellular-based Industrial Automation Systems), in part by Basic Science Research Program through the National Research Foundation of Korea (NRF) funded by the Ministry of Education (No. 2017R1D1A1B03030436), and in part by the NRF funded by the Ministry of Science and ICT (No. 2017R1A5A1015596).

## II. SYSTEM MODEL

The transmit signal $x(m)$ of an OFDM symbol is defined as

$$x(m) = \frac{1}{N} \sum_{k=0}^{N-1} X(k) e^{j2\pi \frac{k}{N}(m-N_{CP})}, \quad (1)$$
$$m = 0, 1, \ldots, N_T - 1$$

where $N$ is the length of the inverse fast Fourier transform (IFFT), $N_{CP}$ is the length of the CP, and $N_T = N + N_{CP}$ denotes the overall length of an OFDM symbol. $X(k)$ denotes the transmit symbol on the $k^{th}$ subcarrier, and $x(m)$ denotes the $m^{th}$ sample of an OFDM symbol. Each OFDM symbol can be shaped by a time waveform (pulse shaping waveform). In this case, the OFDM transmit signal can be written as

$$x(n) = \frac{1}{N} \sum_{l=-\infty}^{\infty} \sum_{k=0}^{N-1} X(k) p(n - lN_T) e^{j2\pi \frac{k}{N}(n-N_{CP}-lN_T)} \quad (2)$$

where the waveform $p(n - lN_T)$ determines the pulse shape. Note that the conventional OFDM signal in (1) corresponds to the case of a rectangular waveform, i.e.,

$$p(m) = \begin{cases} 1, & m = 0, 1, \ldots, N_T \\ 0, & \text{otherwise} \end{cases} \quad (3)$$

The IEEE 802.11af has defined a television very high-throughput (TVHT) physical-layer specifications for the basic channel units (BCUs) of 6, 7, and 8MHz. The sampling clocks are modified to match to each of the BCU bandwidths. Table I shows the major parameters of IEEE 802.11af. It includes 6 MHz, 7 MHz, and 8 MHz bandwidth channels. 6 MHz and 8 MHz bandwidth channels contain 144 sub-carriers, which comprises of 108 data subcarriers, 6 pilot subcarriers, and 30 null subcarriers. A 7 MHz bandwidth channel contains 168 subcarriers, which comprises of 108 data subcarriers, 6 pilot subcarriers, and 54 null subcarriers. Subcarrier indexes corresponding to data subcarriers and pilot subcarriers are as follows:

- Data subcarriers: -58 to -2, 2 to 58
- Pilot carriers: -53, -25, -11, +11, +25, +53

The guard interval is 3 $\mu s$ for 6 MHz and 7 MHz bandwidth channels, and 2.25 $\mu s$ for 8 MHz bandwidth channel. In IEEE 802.11af, the maximum delay spread, i.e., the length of the channel impulse response (CIR) is 1 $\mu s$ [9]. Therefore, guard interval in Table I can prevent OFDM symbols from ISI. In this paper, we will consider the shortest guard interval (and hence possibly the most problematic case) corresponding to the 8MHz BCU to investigate the performance of the proposed filtering method for 802.11af.

## III. PROPOSED PULSE SHAPING SCHEME

This section proposes a novel method to achieve the SEM requirement of 802.11af in the shortest guard interval mode. The proposed method is based on spectrum shaping, using an asymmetric pulse with extended smoothing edge duration, extended guard band, and an FIR filter. Specifically, the proposed method works as the following steps: First, a pulse shaping is employed to shape the spectral leakage below the required power. An asymmetric pulse is adopted to achieve longer smoothing edge duration, which leads to larger spectral leakage suppression while minimizing the ISI. This step will be described in detail in Section III-A. Second, the guard band is extended [2], [10] by increasing the IFFT size to $M$ times the original IFFT size N. The sampling frequency is also increased by $M$ times to maintain the same subcarrier spacing. This step widens the gap between the main spectrum and adjacent image spectra, which allows a short FIR filter to cancel the image spectra. This step is elaborated in Section III-B.

TABLE I. TIMING-RELATED PARAMETERS OF 802.11AF FOR TVHT

| Parameters | 6 MHz | 7 MHz | 8 MHz |
|---|---|---|---|
| Number of data subcarriers | 108 | 108 | 108 |
| Number of pilot subcarriers | 6 | 6 | 6 |
| Total number of subcarriers | 114 | 168 | 114 |
| Highest data subcarrier index | 58 | 58 | 58 |
| Subcarriers frequency spacing | $\frac{6 \text{ MHz}}{144}$ | $\frac{7 \text{ MHz}}{168}$ | $\frac{8 \text{ MHz}}{144}$ |
| IFFT/FFT period | 24 μs | 24 μs | 18 μs |
| Guard interval duration | 3 μs | 3μs | 2.25 μs |

### A. Pulse Shaping

By smoothing the edges of rectangular pulse, the spectral leakage can be reduced [11]. One solution to obtain this is to apply windowing after appending CP and cyclic suffix (CS) before and after each OFDM symbol, respectively, as shown in Fig 1.

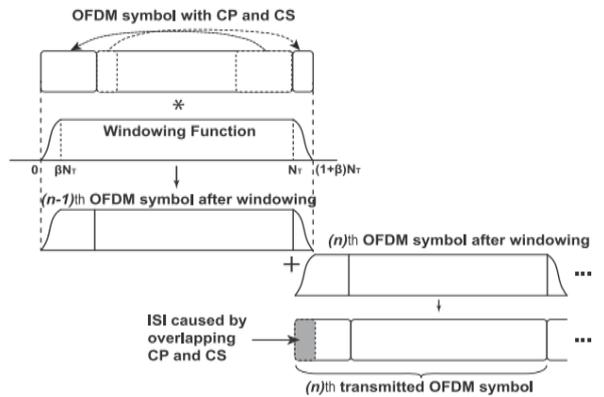

Fig. 1. Pulse shaping operation on OFDM symbols.

In general, an increase in the roll-off factor $\beta$ of the windowing function leads to a decrease in the spectral leakage. However, smoothing edge duration $\beta N_T$ is restricted to be less than the length of the CP to prevent channel-induced ISI [2], [3], [12]. We now investigate three different pulse shapes. The first smoothing function in (4), denoted as $p_1$, is raised cosine function, recommended in the IEEE 802.11a standard:

$$p_1(m) = \begin{cases} \sin^2\left(\frac{\pi}{2}\left(0.5 + \frac{m}{\beta N_T}\right)\right), & -\frac{\beta N_T}{2} \leq m \leq \frac{\beta N_T}{2} \\ 1, & \frac{\beta N_T}{2} < m < N_T - \frac{\beta N_T}{2} \\ \sin^2\left(\frac{\pi}{2}\left(0.5 - \frac{m - N_T}{\beta N_T}\right)\right), & -\frac{\beta N_T}{2} < m < N_T + \frac{\beta N_T}{2} \end{cases} \quad (4)$$

The second, denoted as $p_2$, is based on the characteristics of function with vestigial symmetry as derived in [13]. The function of the vestigial symmetry window is given as

$$p_2(m) = \begin{cases} \frac{1}{2} + \frac{9}{16}\cos\left(\pi\left(1 - \frac{m}{\beta N_T}\right)\right) \\ \quad -\frac{9}{16}\cos\left(\pi\left(1 - \frac{m}{\beta N_T}\right)\right), & 0 \leq m \leq \beta N_T \\ 1, & \beta N_T < m < N_T \\ \frac{1}{2} + \frac{9}{16}\cos\left(\pi\left(\frac{m - N_T}{\beta N_T}\right)\right) \\ \quad -\frac{1}{16}\cos\left(3\pi\left(\frac{m - N_T}{\beta N_T}\right)\right), & N_T < m < (1+\beta)N_T \end{cases} \quad (5)$$

As mentioned in the previous section, asymmetric window can provide a trade-off between spectral efficiency and ISI. This window causes lower ISI in comparison to the symmetric windows [3]. The function of an asymmetric window can be written as

$$p_3(m) = \begin{cases} \cos\left(\frac{\pi(m - \beta N_T)}{2(\beta N_T - 1)}\right), & 0 \leq m \leq \beta N_T \\ 1, & \beta N_T < m < N_T \\ 1 - \cos\left(\frac{\pi(m - \beta N_T)}{2(\beta N_T - 1)}\right), & N_T < m < (1+\beta)N_T \end{cases} \quad (6)$$

Fig. 2 illustrates the shaped spectra using of different pulse shaping windows. *RCn*, *VSn*, and *ASn* denote the shaped spectrum using raised cosines function, vestigial symmetry function and asymmetric pulse, respectively, with the smoothing edge duration equal to the length of *n* original samples. Note that the proposed method employs the asymmetric pulse to obtain a longer smoothing edge duration. *RC4* and *AS4* have almost the same spectral leakage level that is slightly better than *VS4* and significantly better than *RC1*. In the proposed method, the asymmetric pulse is employed to extend the smoothing edge duration while keeping the ISI at acceptable level. As can be seen in Fig. 2, *AS16* denoted the pulse shaping of the proposed method achieves a much better shaped spectrum compared to *RC4*. That gives the possibility to filter the spectrum leakage to meet the SEM requirement of 802.11af. According to the filtering scheme presented in [14] and the parameters of 8MHz channel in Table I, the smoothing edge duration $\beta N_T$ should be conservatively chosen such that it is less than the duration of 4-samples to avoid ISI.

### B. Extending Frequency Guard and FIR Filter

In the above subsection, we do not consider the existence of image spectra, as illustrated in Fig. 4 for the case of up-sampling of 8 times the original frequency, which is caused by the interpolation after the IFFT. Since the guard band of the IEEE 802.11af is narrow, it is necessary to use interpolation to increase sampling frequency and expand the baseband bandwidth. Unfortunately, interpolation causes the appearance of the image spectrum. Because the band gap between the main spectrum and adjacent image spectrum is quite narrow, it is difficult to cancel the image spectra to avoid inter-channel interference (ICI). An FIR filter may be used to eliminate image spectrum [15] [16] [17], caused by interpolation. However, an FIR filter can cause ISI because it reduces the effective guard interval.

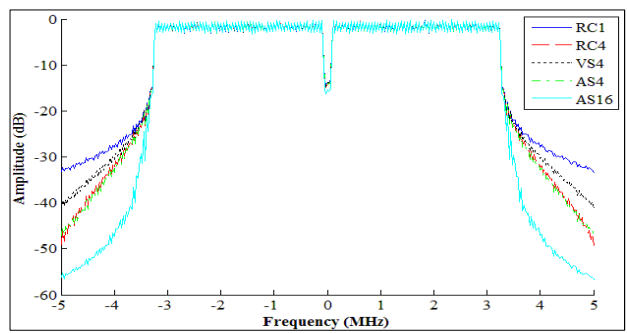

Fig. 2. Comparison of shaped spectra with different shaping schemes.

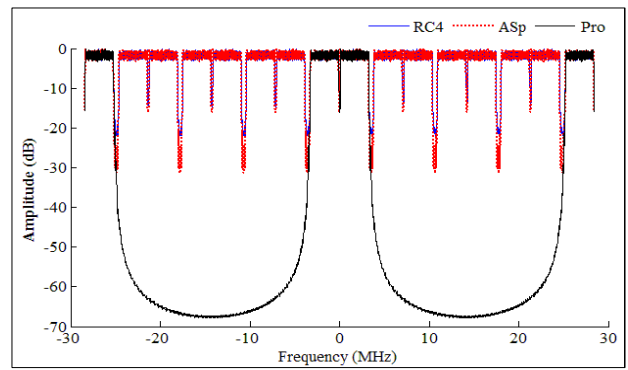

Fig. 3. Comparison of the spectrum after interpolation.

In the proposed method, the guard band is extended by increasing the size of the IFFT to 4 times the original IFFT size, i.e., $N_{FFT} = 4 \times 128 = 512$. As a result, the proposed method needs the interpolation with the factor of 2 instead of 8 to perform up sampling of 8 times. In addition, by doing that the number of image spectra decreases and the gap between the adjacent spectra and the main spectrum increases. This relaxes the requirement of the FIR filter to cancel the image spectra while minimizing the ISI.

Fig. 3 shows the shaped spectrum of the three methods after interpolation. *RC4*, *ASp*, and *Pro* denote the shaped spectrum using raised cosines function with $\beta N_T = 4 * U$, asymmetric pulse function with $\beta N_T = 16 * U$ and the proposed pulse shaping, respectively. The presence of image spectra limits the spectrum leakage shaping of *ASp*. Thanks to the extension of the guard band, *Pro* achieves excellent spectrum shaping.

Table II summarizes the main design parameters of the proposed method as compared to the exiting methods. The CP is chosen from short guard interval mode that equals $16*U$ samples after up sampling. The channel impulse response (CIR) equals *1μs* that is equivalent to 7*U samples after up sampling. The parameters are used for simulation to evaluate the performance of the proposed method the following Section.

TABLE II. PARAMETERS OF THREE APPROACHES

|  | Asymmetric Pulse Shaping | State of The Art Method | Proposed Method |
|---|---|---|---|
| Up Sampling (**U**) | 8 | 8 | 8 |
| NFFT | 128 | 512 | 512 |
| Interpolation | 8 | 2 | 2 |
| Cyclic Prefix | 16*U | | |
| CIR | 7*U | | |
| FIR | 9*U | 5*U | 5*U |
| $\beta N_T$ | 16*U | 4*U | 16*U |

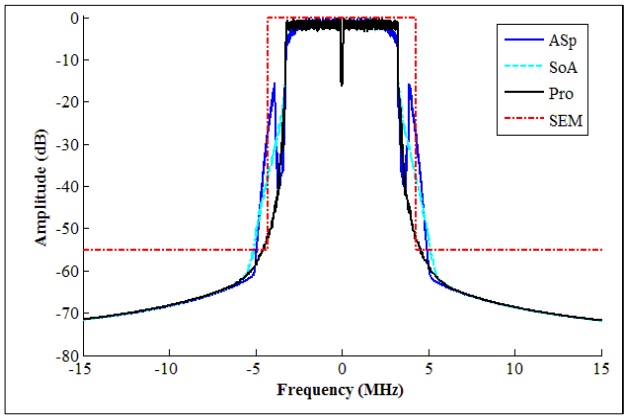

Fig. 4. The filtered spectrum of the proposed method in comparison with the state of the art methods and SEM.

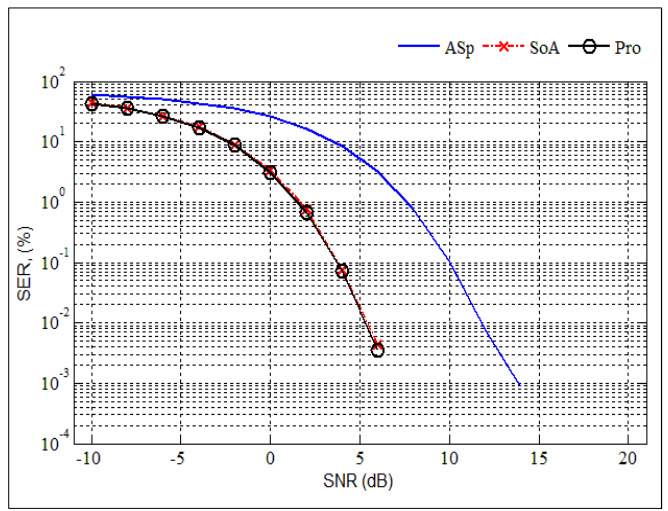

Fig. 5. The comparison of the SER performance in AWGN.

IV. NUMERICAL RESULTS AND DISCUSSION

In this Section, we evaluate the filtering performance of the proposed method in terms of spectrum leakage and SER. Two exiting methods are considered for comparison purpose. One is the conventional filtering based on an asymmetric pulse shaping in [3]. The other is the filtering scheme presented in [14] based on extended guard band. Up sampling by 8 are investigated and the parameters in Table II are used for the three methods.

Fig. 4 illustrates the result of spectrum leakage filtering. SEM represents the Spectral Emission Mask of IEEE 802.11af. As can be seen in Fig. 4, *ASp* has two wide auxiliary peaks beside the main spectrum and introduces visible distortion in the main spectrum. Due to the limited length, the band transition of the FIR filter is not narrow enough to filter out the image spectra. The filtered spectrum of *ASp* method is far from meeting the SEM requirements for 802.11af. The *SoA* method widens the distance between the main spectrum and adjacent image spectra due to extended guard band. This gives the possibility for using a relatively short FIR filter to completely remove the image spectra in the filtered spectrum of *SoA*. However, the *SoA* method uses the symmetric pulse that is constrained with a limited roll-off factor to avoid ISI. Therefore, the side-lobes of *SoA* spectrum are not low enough to meet the SEM requirements. The spectrum leakage of *Pro* is almost as low as the SEM requirements, as the *Pro* uses an asymmetric pulse to lower ISI that allows longer smoothing edge, i.e., larger roll-off factor for pulse shaping. This leads to significant side-slob suppression. In addition, Pro employs extend guard band to cancel the image spectra with a short length FIR filter.

Fig. 5 shows the SEM performance in an AWGN channel. Pro is shown to provide almost identical performance to *SoA*. This means that an increase in the smoothing edge duration of the asymmetric pulse in the proposed method causes negligible ISI. *Pro* achieves better SER than *Asp,* although both of them use the asymmetric pulse shaping. This is because the longer FIR filter in *ASp* causes the larger ISI.

The simulation results show that the proposed method can achieve the spectral leakage suppression to meet the SEM requirement of 802.11af. The ISI caused by performing filtering in the proposed method is negligible, as verified by SER performance.

V. CONCLUSION

This paper presents a novel pulse shaping scheme for efficient spectral leakage suppression in OFDM based physical layer of IEEE 802.11af. The proposed pulse shaping method is based on an asymmetric pulse shaping and extended guard band to overcome the constraint of short guard interval required for high spectral efficiency. The use of asymmetric pulse shaping reduces the effect of ISI leading to the improved SER of the proposed method. The extended guard band allows an efficient suppression of image spectra with a short FIR filter. Numerical results are provided to verify the performance of the proposed scheme in terms of the spectral leakage and the symbol error rate.